\documentclass[amsmath,12pt,amssymb,preprint,prd,aps,nofootinbib]{revtex4}
\usepackage{amsfonts} 
\usepackage{graphicx} 
\usepackage{epsfig}
\usepackage{multirow}
\usepackage{bm}

\begin{document}
\title{Open Charm mesons in magnetized nuclear matter\\
-- effects of (inverse) magnetic catalysis}
\author{Sourodeep De}
\email{sourodeepde2015@gmail.com} 
\author{Amruta Mishra}
\email{amruta@physics.iitd.ac.in}
\affiliation{Department of Physics, Indian Institute of Technology, 
Delhi, Hauz Khas, New Delhi - 110 016, India}

\begin{abstract}
The in-medium masses of the open charm ($D$ and $\bar D$) 
mesons in magnetized isospin asymmetric nuclear matter
are investigated within a chiral effective model,
including the contributions of the Dirac sea (DS) 
to the self-energies of the nucleons.
Within the model, the masses of these mesons are calculated 
from their interactions with the nucleons and the scalar 
($\sigma\sim(\langle\bar u u\rangle+\langle \bar d d \rangle)$, 
$\zeta\sim\langle {\bar s}s\rangle$ and 
$\delta\sim(\langle{\bar u} u\rangle-\langle {\bar d} d\rangle)$)
fields. For zero density, the Dirac sea contributions, 
are observed to lead to an enhancement of the quark condensates 
with increase in magnetic field, an effect called
`magnetic catalysis (MC)'.
This effect is also observed for baryon density, $\rho_B =\rho_0$, 
when anomalous magnetic moments (AMMs) of nucleons are ignored, 
whereas, one observes the opposite trend 
for the quark condensates with the magnetic field, 
i.e., the inverse magnetic catalysis (IMC),
when AMMs are taken into account. 
The Dirac sea contributions are observed to appreciably
modify the open charm meson masses calculated 
within the chiral effective model.
In the presence of an external magnetic field, 
there is mixing of the pseudoscalar and the vector mesons (PV mixing), 
leading to as a drop (increase) in the mass of the pseudoscalar
(longitudinal component of the vector) meson.
The effects of the (inverse) magnetic catalysis, in addition
to the PV mixing and Landau level contribtuions (for
charged mesons) are observed to lead to significant 
modifications to the masses of the open charm mesons.
These can modify the decay widths of the charmonium 
states to open charm mesons, e.g., $\psi(3770)\rightarrow D\bar D$, 
as well as $D^*\rightarrow D\pi$ (and $\bar {D^*}\rightarrow {\bar D}\pi$),
which can affect the production 
of the open charm and charmonium states in ultra 
relativistic peripheral heavy ion collision experiments, 
where the created magnetic field is huge. 

\end{abstract}
\maketitle

\section{Introduction}

The study of the hadron properties under extreme conditions of density, 
temperature and magnetic field is of relevance in the ultra-relativistic 
heavy ion collision experiments and in astrophysical objects like the 
magnetars, neutron stars etc. 
The topic of hadron properties in the presence of magnetic fields 
\cite{Hosaka_Prog_Part_Nucl_Phys} has attracted a lot of attention in the recent past,
due to its relevance in ultra relativistic peripheral heavy ion collision 
experiments, e.g. at RHIC, BNL and LHC, CERN,
where, the magnetic fields produced are estimated to be huge \cite{tuchin}.
The open heavy flavour mesons and the heavy quarkonium states 
are profusely produced in the early phase of these collisions, 
when the magnetic field can still be large. 
However, the time evolution of the magnetic field,
which requires solutions of the magnetohydrodynamic equations,
along with a proper estimate of the electrical conductivity in the 
medium, is still an open question.
The experimental observables of the relativistic heavy ion collision 
experiments are affected by the medium modifications of the hadrons.
In the presence of a magnetic field, there can be an enhancement
(reduction) of the light quark condensate with increase 
in the magnetic field, an effect called the (inverse) 
magnetic catalysis 
\cite{kharzeevmc,kharmc1,chernodub,Preis,menezes,Shovkovy,balicm}.
The (inverse) magnetic catalysis effect has been studied 
for the nuclear matter within Walecka model \cite{haber,arghya}, 
which arises from the Dirac sea contributions to the nucleon 
self energy \cite{haber}. 
In the Walecka model, the effective nucleon mass
is given as $M^*_N=M_N -g_{\sigma N} \sigma$, where 
$M_N$ is the mass of the nucleon in vacuum 
(at zero magnetic field) and $g_{\sigma N}$ is the
coupling of the nucleon with the scalar field, $\sigma$.
The magnetic catalysis (MC) effect is observed 
as an increase in the nucleon mass
with a rise in the magnetic field 
at zero temperature and zero baryon density,
due to the contributions from the Dirac sea
to the self energy of the nucleon \cite{haber,arghya}.
In the no-sea approximation (when the Dirac sea effects
are neglected), the mass of the nucleons remains
at its vacuum value for zero temperature
and zero density. 
Using the Walecka model, the Dirac sea contributions
are calculated through the summation of the
scalar and vector tadpole diagrams in Ref. \cite{arghya}. 
The self energy of the nucleon is calculated
using the weak magnetic field approximation 
for the nucleon propagator, 
and, the effects of the anomalous magnetic moments
(AMM) of the nucleons are also considered. 
For zero temperature and zero density,
there is observed to be quite dominant contributions 
from the AMMs, as compared to when these are not taken 
into account. The AMMs are observed to play
an important role when the Dirac sea contributions
are taken into account. In Ref. \cite{arghya},
at finite densities and zero temperature, 
the AMMs of the nucleons are observed to lead 
to the opposite trend of the drop in the 
effective nucleon mass with the magnetic field, 
and, this trend persists for non-zero  temperatures,
upto the critical temperature, $T_c$. The 
lowering of $T_c$ with increase in the value
of the magnetic field, $B$ is identified with 
the inverse magnetic catalysis (IMC) \cite{arghya}. 
This effect is also seen in lattice studies \cite{balicm}.
In the present work, we calculate the
self-energy of the nucleon 
in magnetized isospin asymmetric nuclear matter
within the chiral effective model,
accounting for the Dirac sea contributions 
to the nucleon self energy
in the weak field approximation for the nucleon
propagator \cite{arghya}, by summing over
the scalar ($\sigma$, $\zeta$ and $\delta$) 
and vector ($\omega$ and $\rho$) tadpole diagrams.

In the presence of a magnetic field, 
there is mixing of the pseudoscalar meson and the 
longitudinal component of the vector meson 
(PV mixing) leading to 
a drop (increase) in the mass of the pseudoscalar 
(longitudinal component of the vector) meson. 
The effects of the PV mixing have been studied 
for the open and hidden charm mesons
\cite{Gubler_D_mag_QSR,charmonium_mag_QSR,charmonium_mag_lee,
Suzuki_Lee_2017,Alford_Strickland_2013}
and have been observed to have appreciable 
modifications to the masses of these mesons.
For the charged mesons, there are additional
contributions from the Landau levels 
in the presence of an external magnetic field. 
The open charm mesons, which are created at the early phase of the heavy ion
collision experiments, when the magnetic field is still large,
can hence be important tools in probing the effects of the magnetic field,
e.g., the mass modifications due to the (inverse) magnetic catalysis,
the mixing of pseudoscalar meson and the longitudinal component 
of the vector meson (PV mixing), additionally, the Landau level 
contributions (for the charged mesons).

The $D(\bar D)$ and $D^*(\bar {D^*})$ mesons are the open heavy 
flavour mesons, which comprise of a heavy charm quark (antiquark) 
and a light (u,d) antiquark (quark). 
Within a chiral effective model, the in-medium masses of the 
$D$ and $\bar D$ mesons have been calculated 
The in-medium properties of the open heavy flavour (charm and bottom) 
mesons amd heavy quarkonium states in the absence of 
a magnetic field have been studied within the model 
\cite{amdmeson,amarindamprc,amarvdmesonTprc,amarvepja,DP_AM_Ds,DP_AM_bbar,DP_AM_Bs,AM_DP_upsilon}.
In the presence of a magnetic field, the masses of the open and hidden
heavy flavour mesons have been studied within the model in the 
`no sea' approximation \cite{dmeson_mag,bmeson_mag,charmonium_mag}
and their effects on the partial decay widths of the charmonium states 
to $D\bar D$ \cite{charmdecay_mag,charmdw_mag,open_charm_mag_AM_SPM} 
and bottomonium states to $B\bar B$ \cite{upslndw_mag}.

In the present work, we 
investigate the masses of the open charm  ($D$ and $\bar D$) mesons
in isospin asymmetric magnetized nuclear matter using a chiral
effective model, accounting for the Dirac sea contributions.
The effects of the AMMs of the nucleons are also considered 
in the present work.
The PV mixing effect is considered using a phenomenological
Lagrangian interaction \cite{charmonium_mag_lee,Iwasaki}.
The additional contributions from the lowest Landau level 
(LLL) are taken into account for the charged open charm mesons.

We organize the paper as follows: In section II, we discuss 
briefly the chiral effective model and the computation of 
the masses of the $D$ and $\bar D$ mesons in the isospin asymmetric
magnetized nuclear matter. These are calculated including the
contributions of the Dirac sea for the nucleon self-energy,
which is observed to a rise (drop) in the magnitude of the 
light quark condensate with increase in the magnetic field, 
an effect called the (inverse) magnetic catalysis.
as well as, incorporating the effects of the PV mixing in the
presence of a magnetic field. In section III, we describe the 
results for the masses of the open charm mesons 
due to the effects of the (inverse) magnetic catalysis, 
PV mixing and for the charged mesons, contributions from
the Landau level. In section IV, we summarize the findings
of the present work. 

\begin{figure}
\vskip -2.5in
    \includegraphics[width=0.9\textwidth]{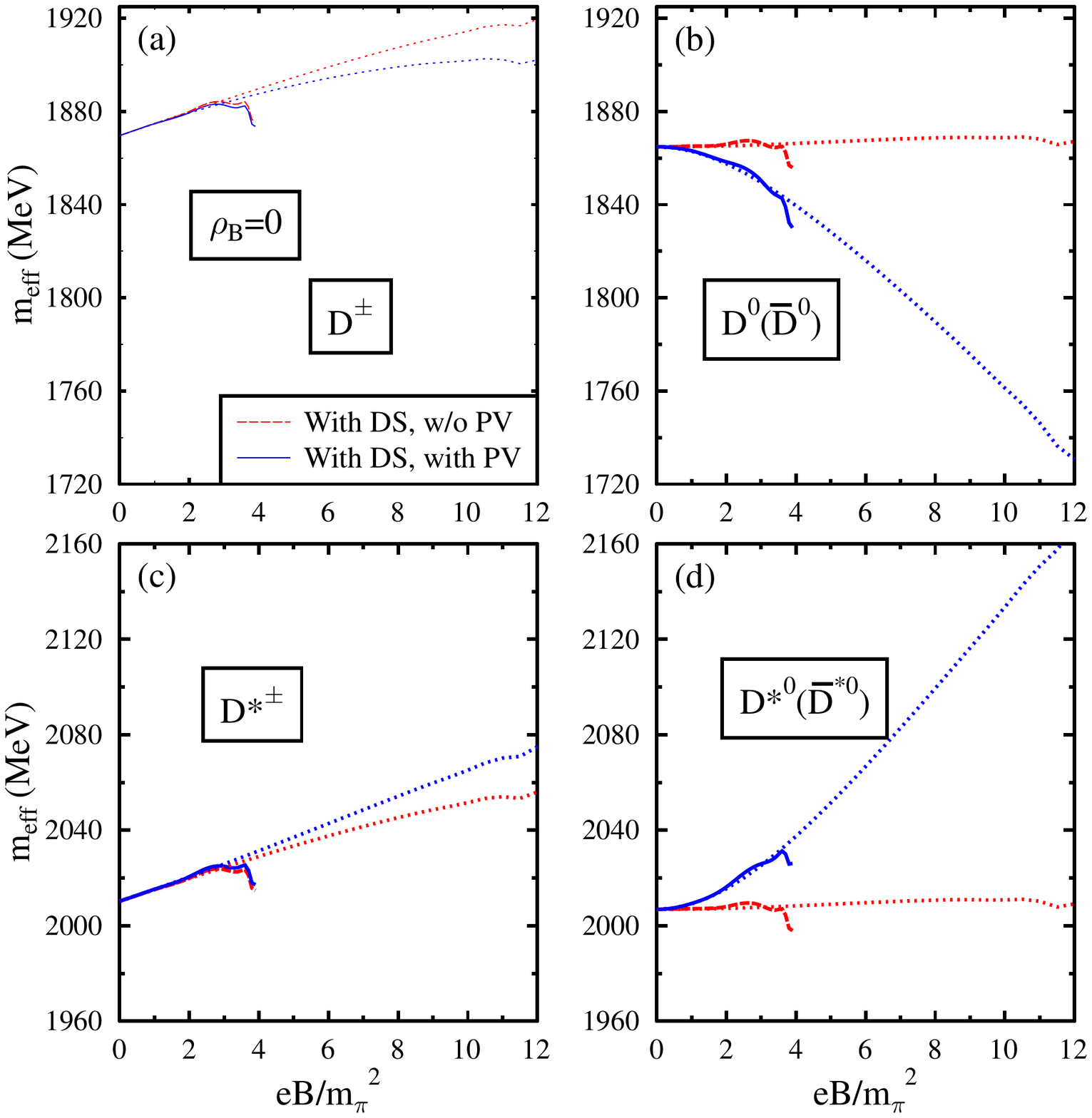}
\vskip -0.3in
     \caption{
     (Color online)
The masses of the $D^{\pm}$, $D^0(\bar {D^0})$, ${D^*}^{\pm}$, 
and ${D^*}^0 (\bar {D^*}^0)$ mesons 
are plotted in (a), (b), (c) and (d) at $\rho_B$=0, including the
contributions of the Dirac sea (DS), and, with and without PV
($D(\bar D)-{D^*}(\bar {D^*})$) mixing effects
accounting for the AMMs of the nucleons. These are compared with 
the cases when the AMMs of the nucleons are not considered
(shown as dotted lines).} 
    \label{mddbar_mag_zero_density_MC}
\end{figure}

\section{In-medium masses of open charm mesons}

We describe the chiral effective model used to study
the open charm $D$ and $\bar D$ mesons in magnetized nuclear matter. 
The model is based on a nonlinear realization of chiral symmetry.
The breaking of scale invariance of QCD is incorporated
into the model through the introduction of a scalar dilaton 
field, $\chi$. The Lagrangian of the model, in the presence of 
a magnetic field, has the form
\begin{equation}
{\cal L} = {\cal L}_{kin} + \sum_ W {\cal L}_{BW}
          +  {\cal L}_{vec} + {\cal L}_0
+ {\cal L}_{scalebreak}+ {\cal L}_{SB}+{\cal L}_{mag}^{B\gamma},
\label{genlag} \end{equation}
where, $ {\cal L}_{kin} $ corresponds to the kinetic energy terms
of the baryons and the mesons,
${\cal L}_{BW}$ contains the interactions of the baryons
with the meson, $W$ (scalar, pseudoscalar, vector, axialvector meson),
$ {\cal L}_{vec} $ describes the dynamical mass
generation of the vector mesons via couplings to the scalar fields
and contains additionally quartic self-interactions of the vector
fields, ${\cal L}_0 $ contains the meson-meson interaction terms,
${\cal L}_{scalebreak}$ is a scale invariance breaking logarithmic
potential given in terms of a scalar dilaton field, $\chi$ and
$ {\cal L}_{SB} $ describes the explicit chiral symmetry
breaking. 
The term ${\cal L}_{mag}^{B\gamma}$, describes the interacion
of the baryons with the electromagnetic field, which includes
a tensorial interaction
$\sim \bar {\psi_i} \sigma^{\mu \nu} F_{\mu \nu} \psi_i$,
whose coefficients account for the anomalous magnetic moments
of the baryons \cite{dmeson_mag,bmeson_mag,charmonium_mag}.

\begin{figure}
\vskip -2.5in
    \centering
    \includegraphics[width=0.9\textwidth]{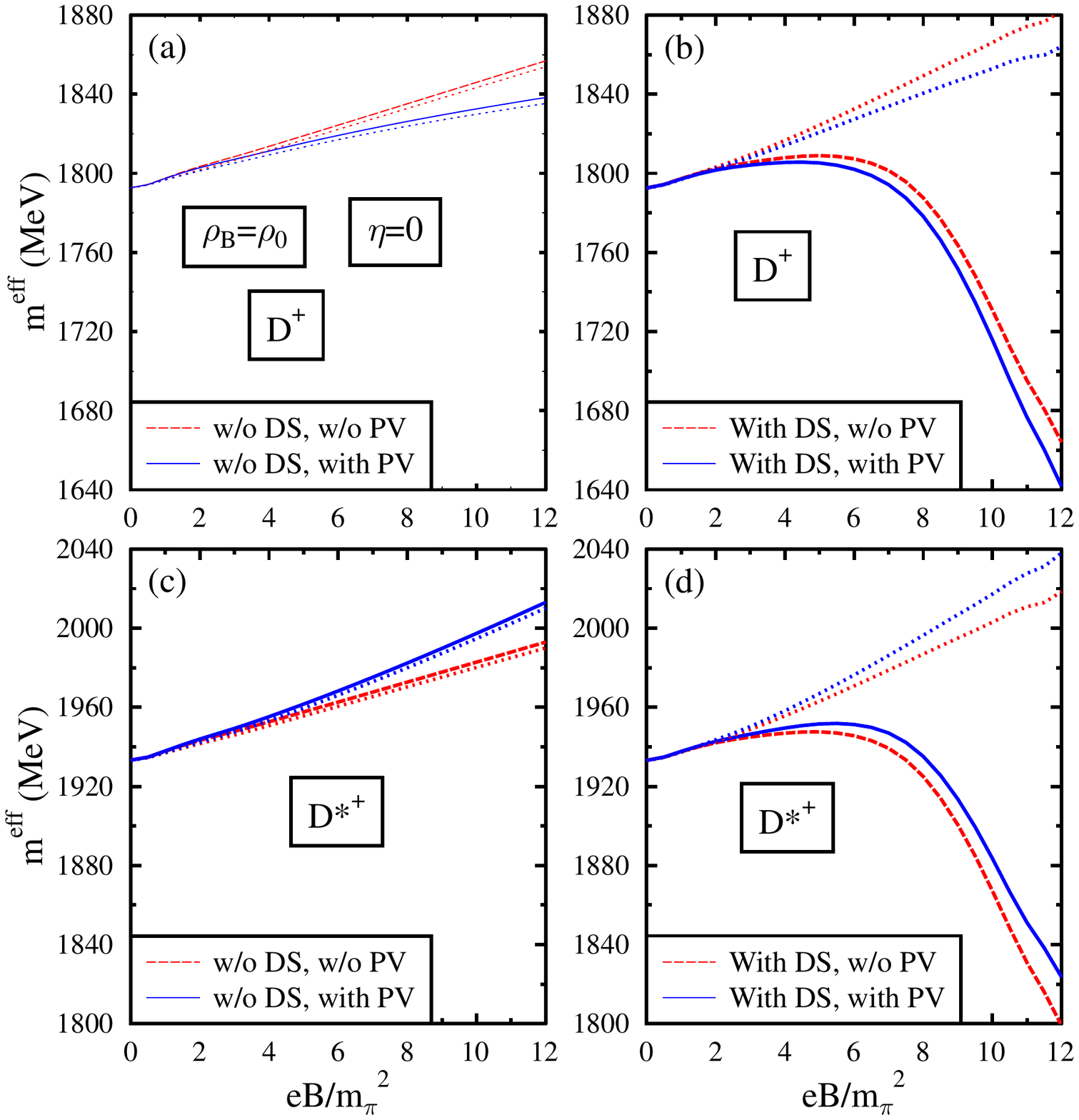}
\vskip -0.3in
     \caption{
     (Color online)
The masses of the $D^+$ and ${{D^*}^+}$ mesons
are plotted for $\rho_B=\rho_0$ in magnetized symmetric
nuclear matter ($\eta$=0)
in (b) and (d) as functions of $eB/m_\pi^2$,
including the
Dirac sea (DS) contributions, and, compared with
the masses obtained without accounting for DS 
(shown in (a) and (c)). The masses are plotted 
with and without PV ($D^+-{{D^*}^+}$ mixing effect, 
accounting for the AMMs of the nucleons. These are compared with 
the cases when the AMMs of the nucleons are not considered
(shown as dotted lines).} 
    \label{mdpdpst_mag_rhb0_eta0_MC}
\end{figure}

\begin{figure}
\vskip -2.5in
    \centering
    \includegraphics[width=0.9\textwidth]{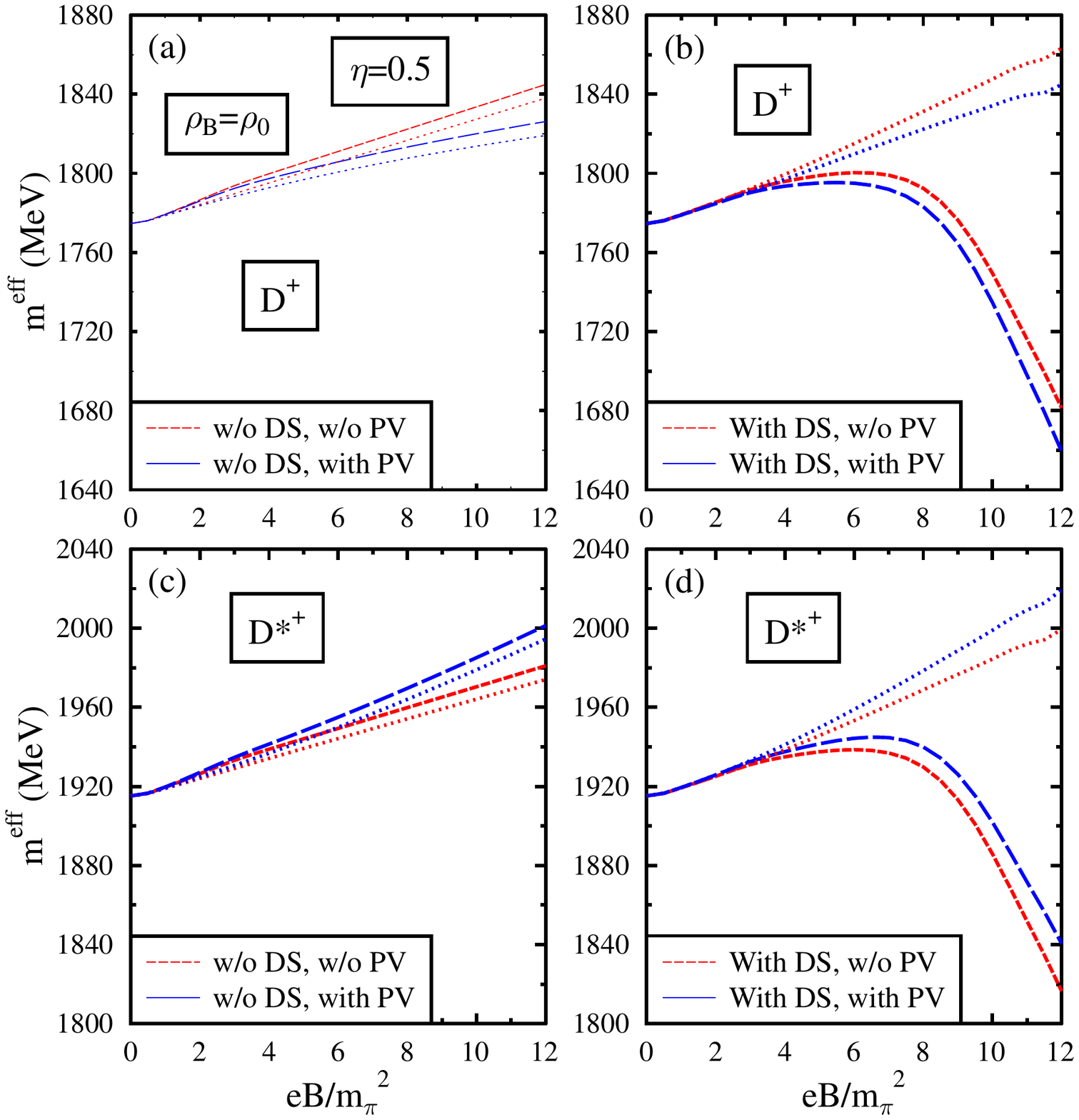}
\vskip -0.3in
     \caption{
(Color online)
     Same as fig \ref{mdpdpst_mag_rhb0_eta0_MC} for $\eta$=0.5.}
    \label{mdpdpst_mag_rhb0_eta5_MC}
\end{figure}

In the present work, the meson fields are treated as claassical 
fields. However, the nucleon is treated as a quantum field and 
the self energy of the nucleon includes the contributions of 
the Dirac sea through the summation of the tadpole diagrams
corresponding to the the non-strange isoscalar ($\sigma$), 
strange isoscalar ($\zeta$), and non-strange isovector 
($\delta$) scalar fields and the vector fields.  
The masses of the baryons are generated by their interactions 
with the scalar mesons. Within the chiral effective model,
the mass of baryon of species $i$
($i=p,n$ in the present work of nuclear matter)
is given as
\begin{equation}
M_i=-g_{\sigma i}\sigma -g_{\zeta i}\zeta -g_{\delta i}\delta, 
\end{equation}
where the scalar fields $\sigma$, $\zeta$ and $\delta$ 
are solved along with the dilaton field, $\chi$, from
ther coupled equations of motion. The the scalar 
densities of the proton and neutron, occurring in the
equations of the scalar fields ($\sigma$, $\zeta$ and 
$\delta$) include the contributions from the Dirac sea,
calculated in the weak field approximation for the 
nucleon propagator \cite{arghya}.  
The masses of the $D$ and $\bar D$ mesons are computed 
from solution of their dispersion relations, given as
\cite{dmeson_mag}
\begin{equation}
-\omega^2+ {\vec k}^2 + m_{D(\bar D)}^2
 -\Pi_{D(\bar D)}(\omega, |\vec k|)=0,
\label{dispddbar}
\end{equation}
where $\Pi_{D(\bar D)}$ denotes the self energy
of the $D$ ($\bar D$) meson in the medium.
For the $D$ meson doublet ($D^0$,$D^+$), and $\bar D$ meson
doublet (${\bar D}^0$,$D^-$), the self enrgies are given by
\begin{eqnarray}
\Pi (\omega, |\vec k|) &= & \frac {1}{4 f_D^2}\Big [3 (\rho_p +\rho_n)
\pm (\rho_p -\rho_n) \big)
\Big ] \omega \nonumber \\
&+&\frac {m_D^2}{2 f_D} (\sigma ' +\sqrt 2 {\zeta_c} ' \pm \delta ')
\nonumber \\ & +& \Big [- \frac {1}{f_D}
(\sigma ' +\sqrt 2 {\zeta_c} ' \pm \delta ')
+\frac {d_1}{2 f_D ^2} (\rho^s_p +\rho^s_n)\nonumber \\
&+&\frac {d_2}{4 f_D ^2} \Big (({\rho^s_p} +{\rho^s_n})
\pm   ({\rho^s_p} -{\rho^s_n}) \Big ) \Big ]
(\omega ^2 - {\vec k}^2),
\label{selfd}
\end{eqnarray}
and
\begin{eqnarray}
\Pi (\omega, |\vec k|) &= & -\frac {1}{4 f_D^2}\Big [3 (\rho_p +\rho_n)
\pm (\rho_p -\rho_n) \Big ] \omega\nonumber \\
&+&\frac {m_D^2}{2 f_D} (\sigma ' +\sqrt 2 {\zeta_c} ' \pm \delta ')
\nonumber \\ & +& \Big [- \frac {1}{f_D}
(\sigma ' +\sqrt 2 {\zeta_c} ' \pm \delta ')
+\frac {d_1}{2 f_D ^2} (\rho^s_p +\rho^s_n
)\nonumber \\
&+&\frac {d_2}{4 f_D ^2} \Big (({\rho^s_p} +{\rho^s _n})
\pm   ({\rho^s_n} -{\rho^s_n}) \Big ]
(\omega ^2 - {\vec k}^2),
\label{selfdbar}
\end{eqnarray}
where the $\pm$ signs refer to the $D^0$ and $D^+$ respectively
in equation (\ref{selfd}) and
to the $\bar {D^0}$ and $D^-$ respectively in equation (\ref{selfdbar}).
In equations (\ref{selfd}) and (\ref{selfdbar}),
$\sigma'(=(\sigma-\sigma _0))$,
${\zeta_c}'(=(\zeta_c-{\zeta_c}_0))$ and  $\delta'(=(\delta-\delta_0))$
are the fluctuations of the scalar-isoscalar fields $\sigma$ and $\zeta_c$,
and the third component of the scalar-isovector field, $\delta$,
from their vacuum expectation values (for zero magnetic field).
 
The values of the scalar meson fields $\sigma$, $\zeta$, $\delta$  
are obtained by solving their coupled equations of motion.
In the present work, as has already been mentioned,
the contribution of the Dirac sea
to the self energy of the nucleon is taken into account
through the summation of the tadpole diagrams,
which are incorporated into the equations of motion
of the fields $\sigma$, $\zeta$ and $\delta$ through
the scalar densities of the nucleons.
These scalar fields, along with the dilaton field, $\chi$, 
are calculated for a given 
baryon density, $\rho_B$, given isospin asymmetry,
$\eta=(\rho_n-\rho_p)/(2\rho_B)$, where $\rho_{p,n}$
are the number densities of the proton and neutron
respectively. Accounting for the lowest Landau level (LLL)
contributions for the charged open charm mesons,
the effective mass of $D^\pm$ is given as
\begin{equation}
m^{eff}_{D^\pm}=\sqrt {{m^*_{D^\pm}}^2 +|eB|},
\label{mdpm_landau}
\end{equation}
whereas for the neutral ($D^0$ and $\bar {D^0}$) mesons,
the effective masses are given as
\begin{equation}
m^{eff}_{D^0 (\bar {D^0})}=m^*_{D^0 (\bar {D^0})}.
\label{md0d0bar}
\end{equation}
In equations (\ref{mdpm_landau}) and (\ref{md0d0bar}),
$m^*_{D^\pm,D^0,\bar {D^0}}$ are the masses calculated
using the chiral effective model, as the solutions for
$\omega$ at $|\vec k|=0$, of the dispersion relations
for these mesons given by equation (\ref{dispddbar}).

\begin{figure}
\vskip -2.5in
    \centering
    \includegraphics[width=0.9\textwidth]{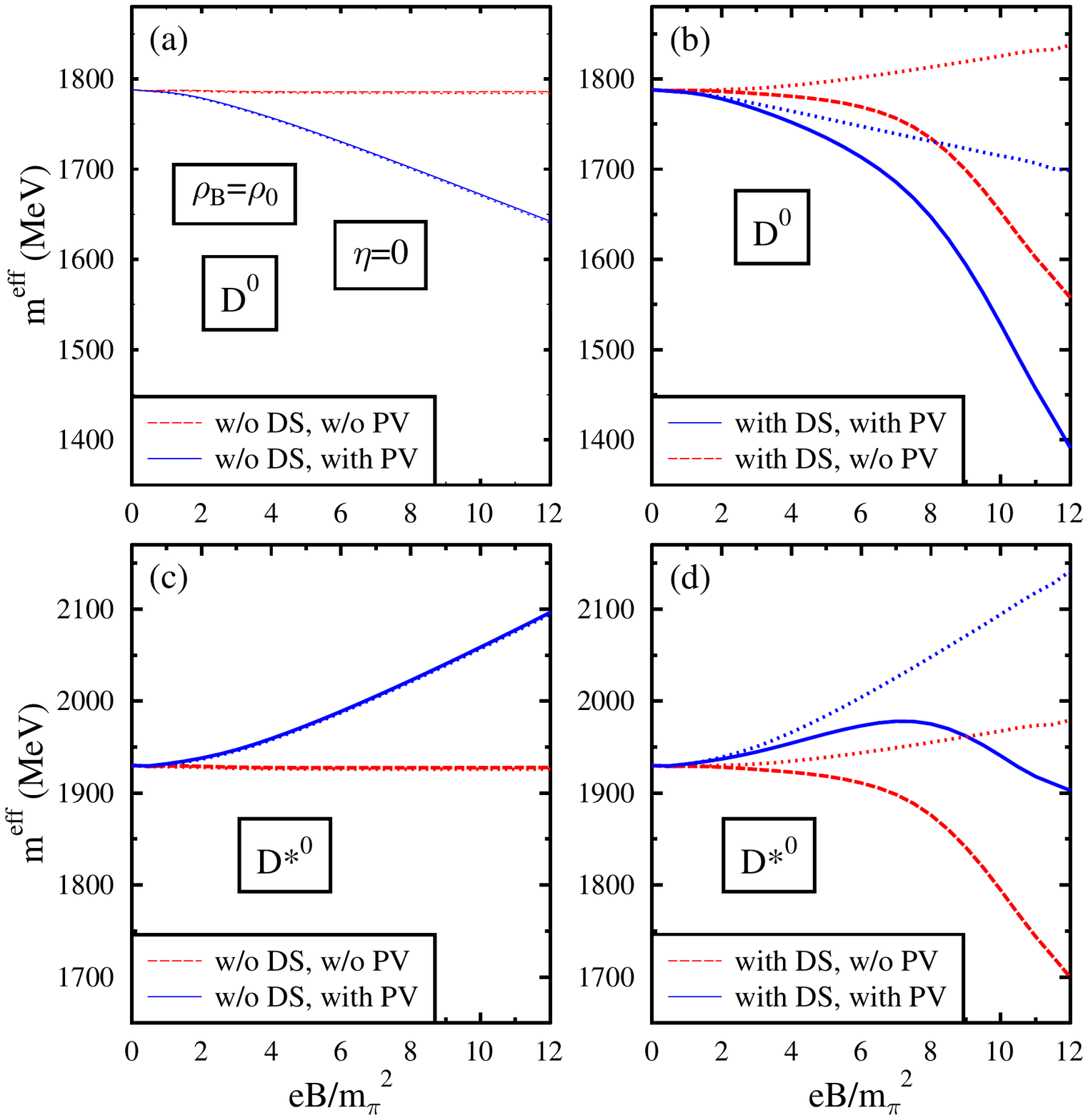}
\vskip -0.3in
     \caption{
     (Color online)
The masses of the $D^0$ and ${{D^*}^0}$ mesons
are plotted for $\rho_B=\rho_0$ in magnetized symmetric
nuclear matter ($\eta$=0) in (b) and (d) as functions of $eB/m_\pi^2$,
including the
Dirac sea (DS) contributions, and, compared with
the masses obtained without accounting for DS 
(shown in (a) and (c)). The masses are plotted 
with and without PV ($D^0-{{D^*}^0}$) mixing effect, 
accounting for the AMMs of the nucleons. These are compared with 
the cases when the AMMs of the nucleons are not considered
(shown as dotted lines).} 
    \label{md0d0st_mag_rhb0_eta0_MC}
\end{figure}

\begin{figure}
\vskip -2.5in
    \centering
    \includegraphics[width=0.9\textwidth]{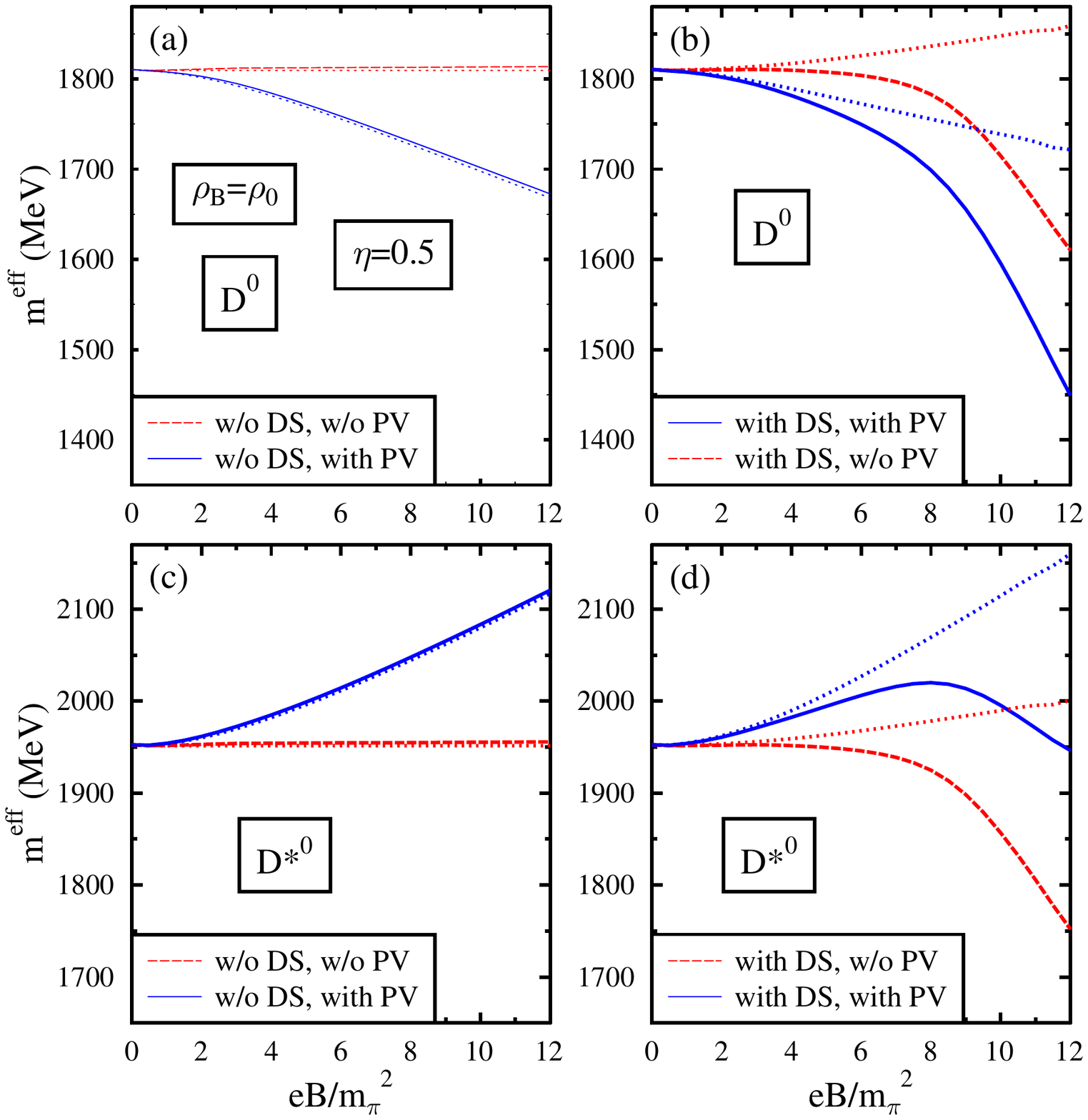}
\vskip -0.3in
     \caption{
(Color online)
     same as fig \ref{md0d0st_mag_rhb0_eta0_MC} for $\eta$=0.5.}
    \label{md0d0st_mag_rhb0_eta5_MC}
\end{figure}

\begin{figure}
\vskip -2.5in
    \centering
    \includegraphics[width=0.9\textwidth]{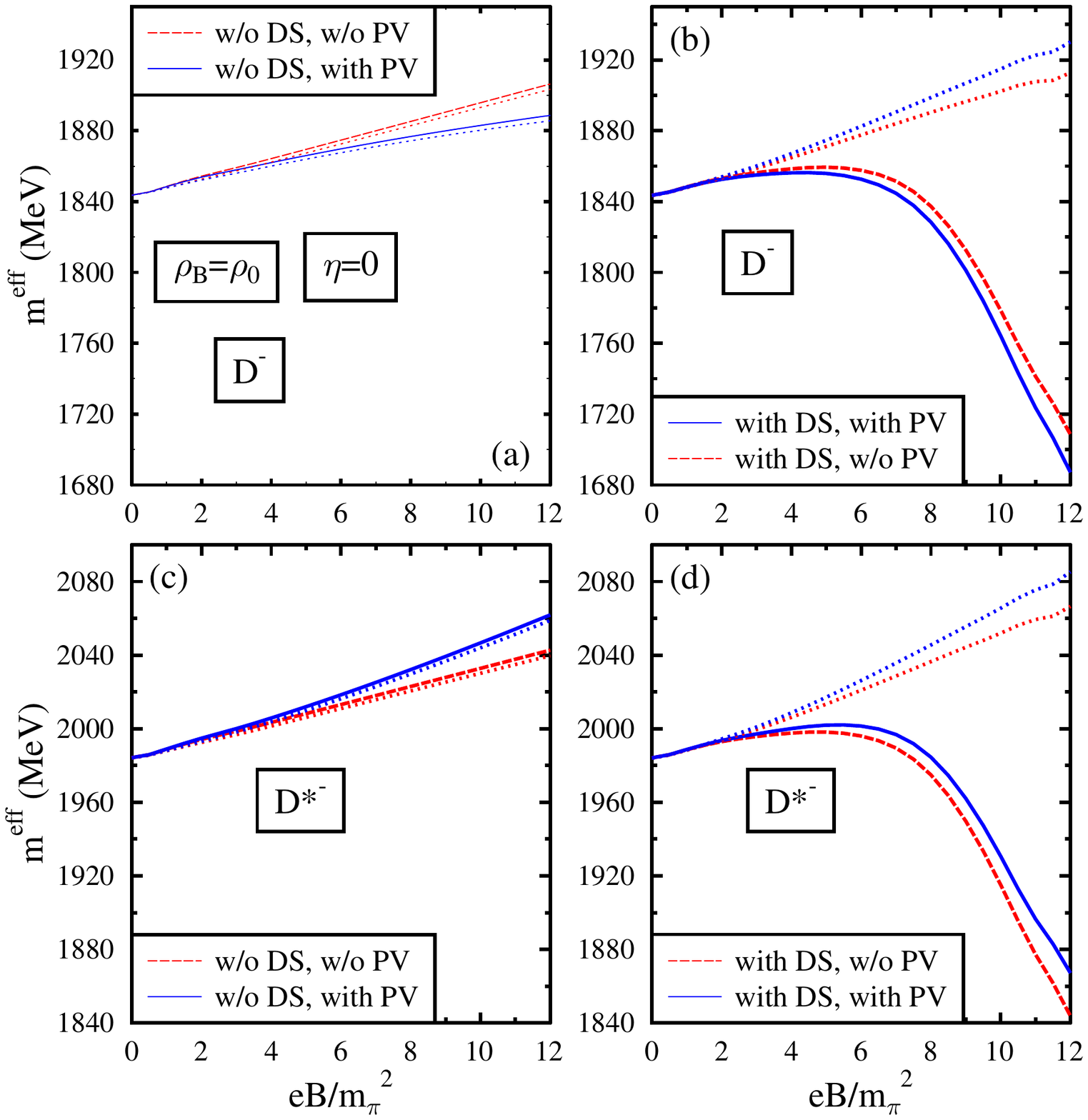}
\vskip -0.3in
     \caption{
     (Color online)
The masses of the $D^-$ and ${{D^*}^-}$ mesons
are plotted for $\rho_B=\rho_0$ in magnetized symmetric
nuclear matter ($\eta$=0) in (b) and (d) as functions of $eB/m_\pi^2$,
including the
Dirac sea (DS) contributions, and, compared with
the masses obtained without accounting for DS 
(shown in (a) and (c)). The masses are plotted 
with and without PV ($D^--{{D^*}^-}$) mixing effect, 
accounting for the AMMs of the nucleons. These are compared with 
the cases when the AMMs of the nucleons are not considered
(shown as dotted lines).} 
    \label{mdmdmst_mag_rhb0_eta0_MC}
\end{figure}

\begin{figure}
\vskip -2.5in
    \centering
    \includegraphics[width=0.9\textwidth]{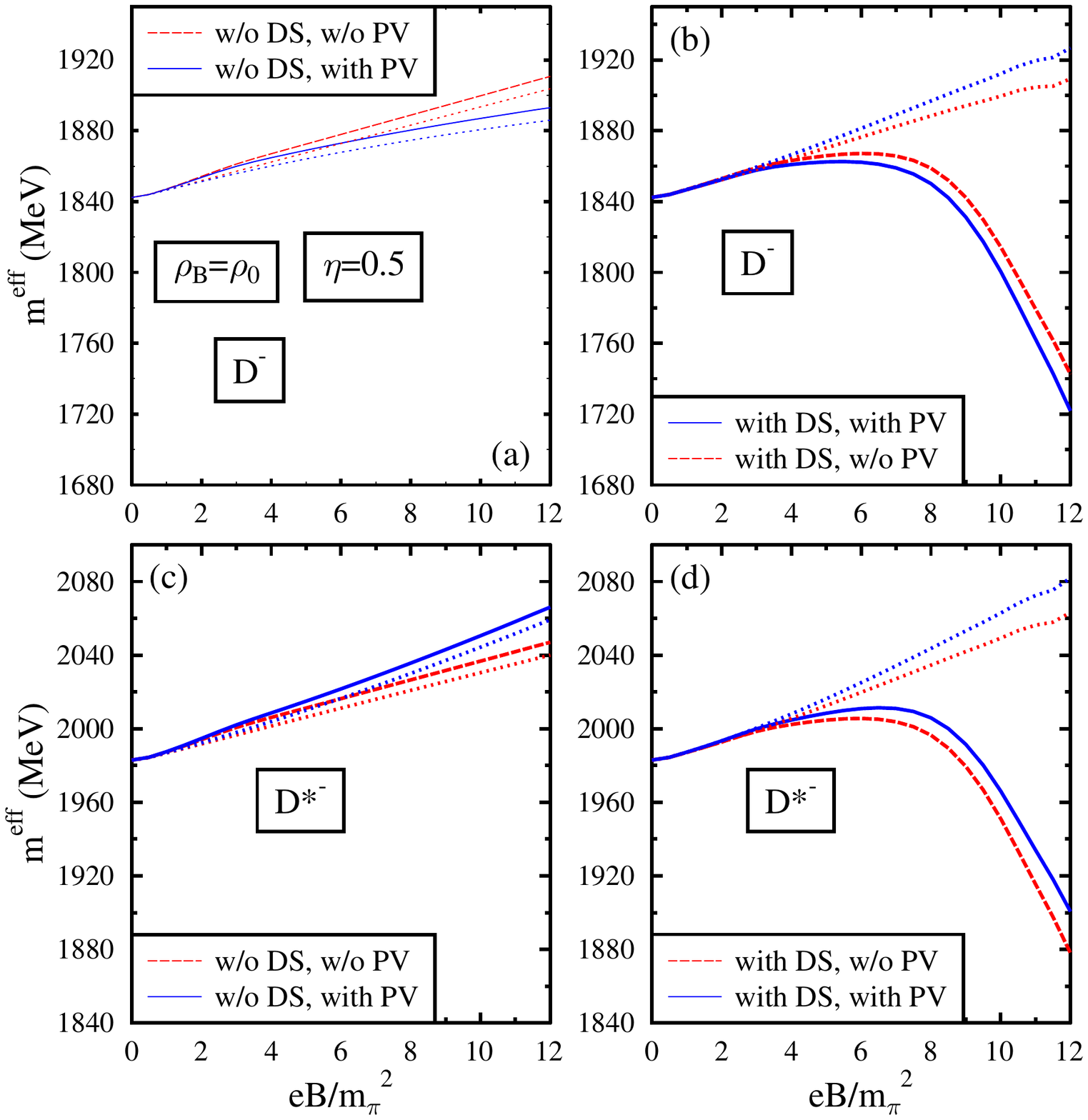}
\vskip -0.3in
     \caption{
(Color online)
     Same as fig \ref{mdmdmst_mag_rhb0_eta0_MC} for $\eta$=0.5.}
    \label{mdmdmst_mag_rhb0_eta5_MC}
\end{figure}

The masses of the charged vector open charm 
($D^*$ and $\bar {D^*}$) mesons, 
retaining the lowest Landau level contributions ($n$=0), 
depend on $S_z$ (the z-component of the spin vector) 
and are given as
\begin{equation}
m^{eff}_{{D^*}^\pm}=\sqrt {{m^*}_{{D^*}^\pm}^2+|eB|
+gS_z|eB|},
\label{mdpmstr_landau}
\end{equation}
whereas for the neutral ${D^*}^0$ and $\bar {{D^*}^0}$,
the in-medium masses are given as
\begin{equation}
m^{eff}_{D^*(\bar {D^*})}={m^*}_{D^*(\bar {D^*})}.
\label{md0std0stbar}
\end{equation}
It is assumed that the mass shifts of the vector open charm 
($D^*$ and $\bar {D^*}$) mesons (which have same quark-antiquark
constituents as $D$ and $\bar D$ mesons) are identical to the mass shifts
of the pseudoscalar mesons $D$ and $\bar D$ mesons, calculated 
within the chiral effective model \cite{open_charm_mag_AM_SPM}.
This is in line with the QMC model where the masses of the hadrons
are obtained from the modification of the scalar density 
of the light quark (antiquark) constituent of the hadron
\cite{Hosaka_Prog_Part_Nucl_Phys}. The in-medium mass
of the vector open charm mesons are thus assumed to be
\begin{equation}
    m^*_{D^*(\bar {D^*})}-m_{D^*(\bar {D^*})}^{vac} 
= {m^*}_{D(\bar D)} - m_{D(\bar D)}^{vac},
\label{mdstrdbatstr}
\end{equation}
for which there are additional Landau level contributions 
for the charged ${D^*}^\pm$ mesons, as given by equation
(\ref{mdpmstr_landau}). In the presence of an external
magnetic field, there is also 
mixing of the pseudoscalar and the longitudinal component ($S_z$=0) 
of the vector meson. Its effect on the masses of the
open charm mesons is also studied in the present work.

\begin{figure}
\vskip -2.5in
    \centering
    \includegraphics[width=0.9\textwidth]{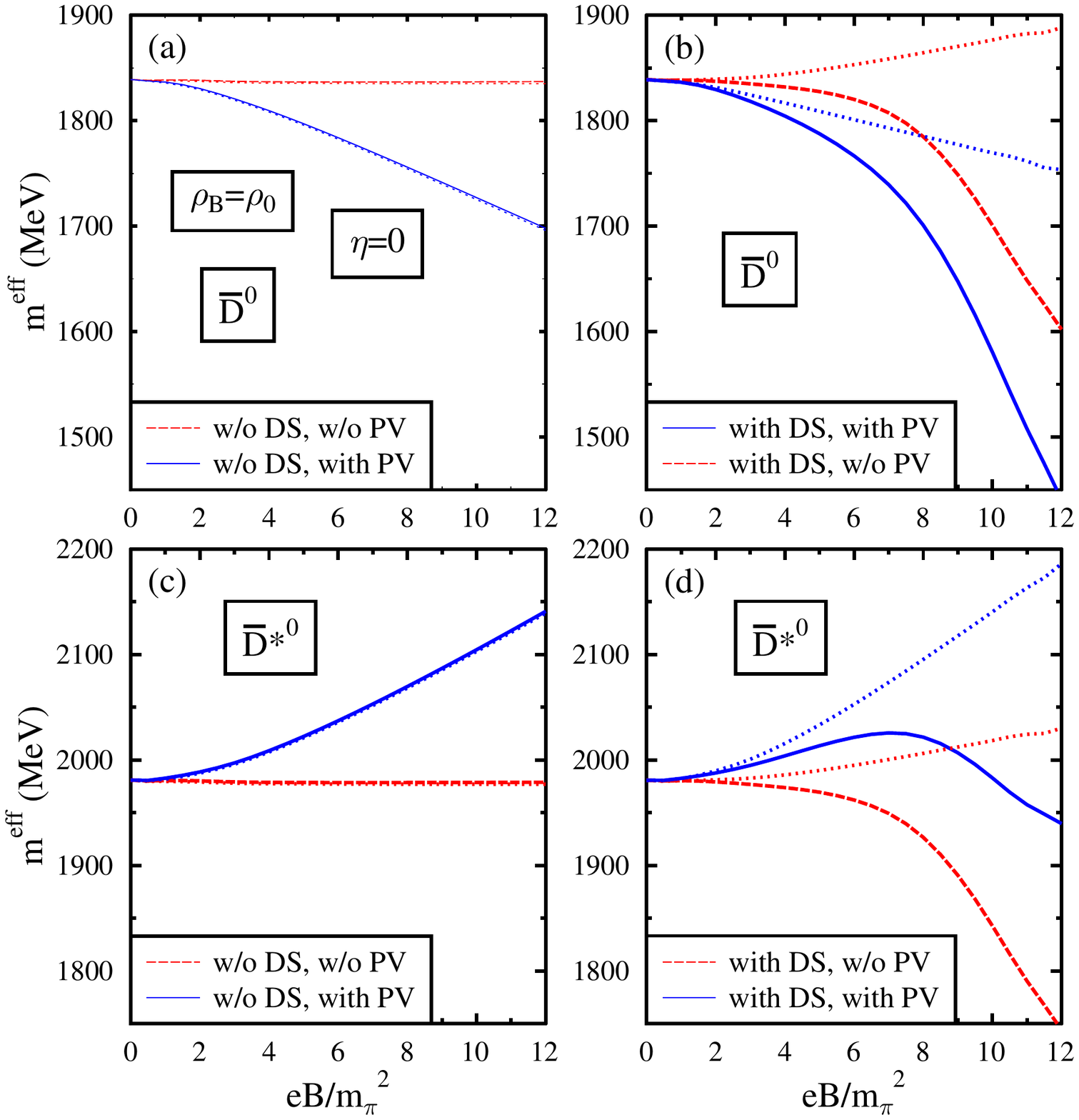}
\vskip -0.3in
     \caption{
     (Color online)
The masses of the $\bar {D^0}$ and $\bar {{D^*}^0}$ mesons
are plotted for $\rho_B=\rho_0$ in magnetized symmetric
nuclear matter ($\eta$=0) in (b) and (d) as functions of $eB/m_\pi^2$,
including the
magnetic catalysis (MC) effect, and, compared with
the masses when without accounting for MC effect
(shown in (a) and (c)). The masses are plotted 
with and without PV ($\bar {D^0}-\bar {{D^*}^0}$) mixing effect, 
accounting for the AMMs of the nucleons. These are compared with 
the cases when the AMMs of the nucleons are not considered
(shown as dotted lines).} 
    \label{md0bd0bst_mag_rhb0_eta0_MC}
\end{figure}

\begin{figure}
\vskip -2.5in
    \centering
    \includegraphics[width=0.9\textwidth]{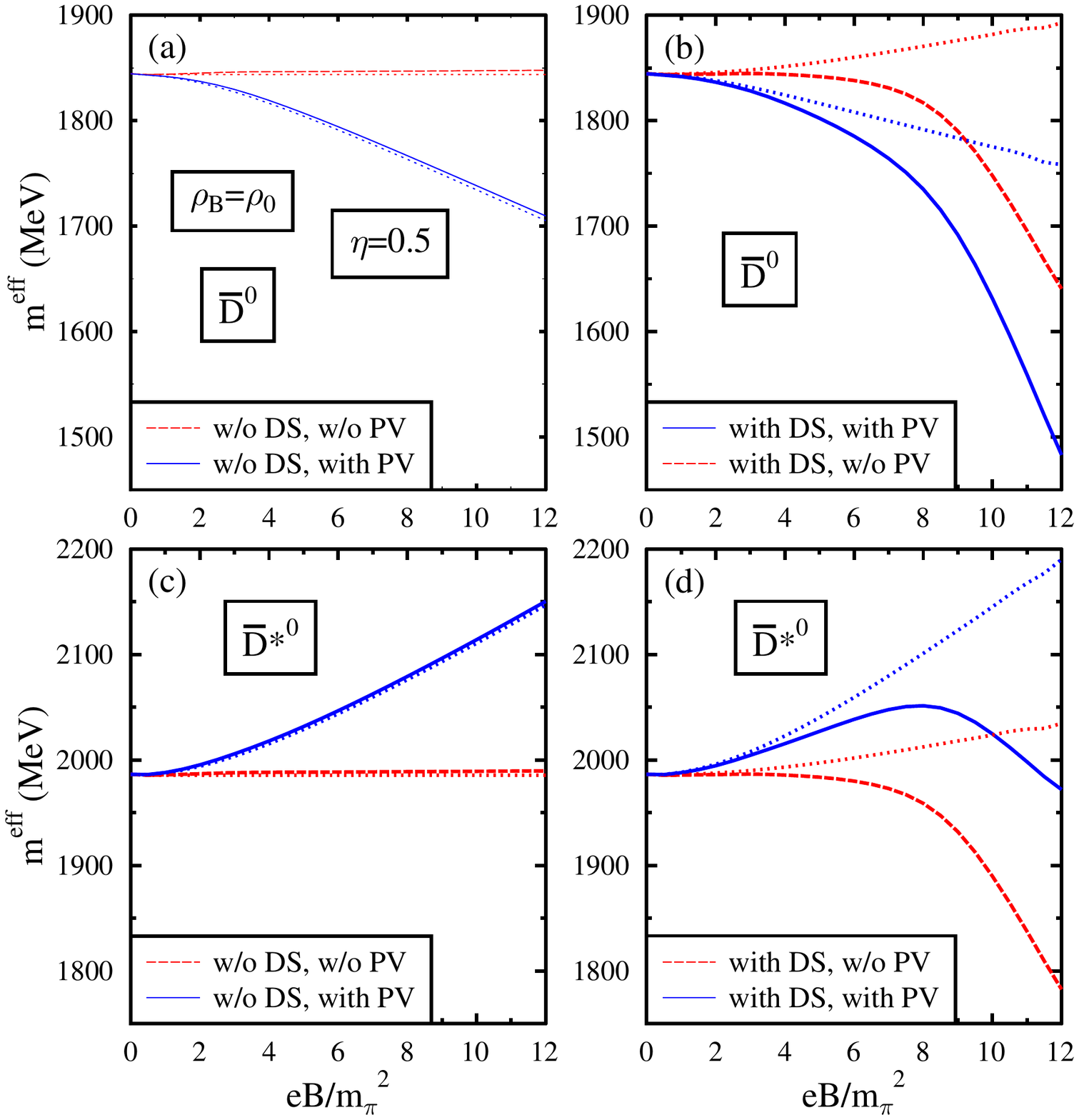}
\vskip -0.3in
     \caption{
(Color online)
     Same as fig \ref{md0bd0bst_mag_rhb0_eta0_MC} for $\eta$=0.5.}
    \label{md0bd0bst_mag_rhb0_eta5_MC}
\end{figure}

The mixing between the pseudoscalar
and the longitudinal component of the vector (PV mixing) mesons
in the presence of a magnetic field, 
is observed to lead to appreciable modifications to their masses
\cite{charmonium_mag_QSR,Gubler_D_mag_QSR,charmonium_mag_lee,Suzuki_Lee_2017,Alford_Strickland_2013,charmdw_mag,open_charm_mag_AM_SPM,upslndw_mag,Iwasaki}.
In the present work of the study of the in-medium masses
of the open charm mesons, the PV mixing effects for the neutral
open charm mesons ($D^0-{{D^*}^0}^{||}$ and $\bar {D^0}-\bar {{D^*}^0}^{||}$
mixings) as well as the charged mesons
($D^+-{{D^*}^+}^{||}$ and ${D^-}-{{D^*}^-}^{||}$ mixings) are considered.
The PV mixing is taken into account through a phenomenological 
Lagrangian which was used to study the mixing of the charmonium states 
\cite{charmonium_mag_lee,Suzuki_Lee_2017}. 
The effects due to the PV ($J/\psi-\eta_c$, 
$\psi(2S)-\eta_c(2S)$ and $\psi(1D)-\eta_c(2S)$) 
mixings on the masses of the charmonium states
in magnetized nuclear matter calculated within a chiral
effective model, have been studied in Ref. \cite{charmdw_mag}.
The effects of the mass modifications due to PV mixing
on the decay width $\psi(3770)\rightarrow D\bar D$
were observed to be quite appreciable due to the
$\psi(1D)-\eta_c(2S)$ mixing \cite{charmdw_mag}, 
as well as due to the PV mixing of the open charm
($D(\bar D)-D^*(\bar {D^*})$
 mesons \cite{open_charm_mag_AM_SPM}.
The $PV\gamma$ interaction Lagrangian is given as,
 \begin{equation}
     \mathcal{L}_{PV\gamma}=\frac{g_{PV\gamma}}{m_{av}}
e\tilde{F}_{\mu\nu}(\partial^{\mu} P)V^{\nu},
 \end{equation}
where P and $V^\mu$ represent the pseudoscalar and the vector fields, 
respectively, $\tilde{F}_{\mu\nu}$ is the dual field 
strength tensor of the external magnetic field and
and $m_{av}$ is the average of the masses of the pseudoscalar 
and vector mesons, $(m_{av} = (m_P+m_V)/2)$. 
$g_{PV}$ is the coupling constant for the radiative decay,
$V\rightarrow P \gamma$. The value of the coupling parameter,
$g_{PV\gamma}$ is fixed from the observed decay width
of $V\rightarrow P\gamma$.
The masses of the pesudoscalar and the longitudinal 
component of the vector meson, due to the PV mixing
are given as
\begin{equation}
     m_{V^{||},P}^2=M_+^2+\frac{c_{PV}^2}{m_{av}^2}
\pm\sqrt{M_-^4+\frac{2c_{PV}^2 M_{+}^2}{m_{av}^2}
+\frac{c_{PV}^4}{m_{av}^4}} 
 \end{equation}
  with $M_{\pm}^2 = m_V^{{eff}^2} \pm m_P^{{eff}^2}$ 
and $c_{PV}=g_{PV\gamma}eB$; with $m_{V,P}^{eff}$ are 
the effective masses of the vector and pseudoscalar 
mesons.These effective masses, given by
(\ref{mdpm_landau}), (\ref{md0d0bar}),
(\ref{mdpmstr_landau}) and (\ref{md0std0stbar}),
include the effects of the Dirac sea calculated 
in the chiral effective model, with additional 
contributions due to the lowest Landau level 
contributions for the charged mesons.
 
\section{Results and Discussions}

In the present work, we study the effects of magnetic field
on the pseudoscalar ($D$ and $\bar D$) and vector ($D^*$ and 
$\bar {D^*}$) open charm mesons in magnetized (nuclear) matter. 
The in-medium masses are calculated using a chiral effective 
model for the $D$ and $\bar D$ mesons due to their interactions
with the nucleons and the scalar mesons, including the
Dirac sea contributions, with additional
Landau level contributions for the charged $D^\pm$ mesons. 
It might be noted
here the neutron, which is electrically charge neutral, interacts
with the magnetic field only due to its non-zero value of the
anomalous magnetic moment.
The effects of the magnetic field
on the masses are computed for zero baryon density as well as, 
for $\rho_B=\rho_0$, the nuclear matter saturation density,
The effects of isospin asymmetry as well as anomalous magnetic
moments (AMMs) of the nucleons are also studied 
in the present work. 
The Dirac sea contributions lead to increase in the values of the
scalar fields, $\sigma (\sim \langle \bar u u\rangle +
\langle \bar d d\rangle)$ and $\zeta (\sim \langle \bar s s\rangle)$,
leading to decrease in the nucleon self energy, i.e., enhancement
of the light quark condensates (an effect called magnetic catalysis),
as compared to when this effect is not considered. 
The values (in MeV) of $\sigma$ and $\zeta$ are modified from the zero
baryon density and zero magnetic fields values of $-93.3$ and $-106.6$
to $-94.537 (-102.018) $ and $-107.19 (-109.9)$ for $eB=3 m_\pi^2$,
without (with) accounting for the AMMs of the nucleons.
These lead to the effective nucleon masses to be
948 (1026) MeV in the absence (presence) of the AMMs of the
nucleons. Hence there is a dominant rise in the
nucleon mass due to the AMMs of the nucleons,
as compared to when AMMs are not taken into 
account, which was also observed in Ref. \cite{arghya}.

In figure \ref{mddbar_mag_zero_density_MC}, at zero baryon density,
the masses of the charged and neutral pseudoscalar and vector 
mesons are plotted accounting for the Dirac sea contributions
to the nucleon self-energy (observed to lead to magnetic catalysis)
as well as PV mixing effects. The $D-{D^*}$ and $\bar D- \bar{D^*}$)
mixings result in a drop (increase) in the mass of the $D({D^*}^{||})$
and $\bar D(\bar {D^*}^{||})$ meson.
The open charm masses are plotted accounting for the AMMs of the nucleons 
and compared to the cases when the AMMs are not taken into account
(shown as dotted lines). 
While accounting for AMMs of the nucleons, at zero density,
the solutions for the scalar fields do not exist
for $eB$ higher than around $4 m_\pi^2$
in the weak field approximations as used here.
The masses for $D^\pm$ are observed to be identical 
to each other (so also for the $D^0$ and $\bar {D^0}$ mesons).
This can be understood from 
the disperion relations of the $D$ and $\bar D$ mesons
given by (\ref{dispddbar}), (\ref{selfd}) and (\ref{selfdbar}).
For $\rho_B=0$ (hence $\rho_p$ and $\rho_n$ are both zero), 
implying that the Weinberg-Tomozawa contribution is zero.
The dispersion relation satisfied by the $D^+(D^0)$ meson is then
same as that of $D^-(\bar {D^0})$ meson, and hence, the masses
are identical within the charged and the neutral sectors.
The effective masses of ${D^*}(\bar {D^*})$ mesons, 
which are calculated, using equation (\ref{mdstrdbatstr}) 
are also identical within the charged as well as neutral sectors. 
The effects of the AMMs on the $D$ and $\bar D$ mesons
are through the Dirac sea (DS)
contributions to the scalar densities of proton and
neutron, $\rho_p^s$ and $\rho_n^s$, occurring in the
$d_1$ and $d_2$ range terms in their self-energies, 
as well as, due to the 
values of the scalar fields, which are calculated accounting
for contributions from Dirac sea through $\rho_p^s$ and 
$\rho_n^n$. The modifications due to AMMs are observed 
to be negligible upto $eB\sim 2 m_\pi^2$ and remains
small for higher values of the magnetic field,
upto $eB\sim 4 m_\pi^2$, beyond which the solutions 
of the scalar fields do not exist for $\rho_B=0$,
when AMMs of nucleons are taken into account,
in the weak magnetic field approximation
\cite{arghya} as used in the present work.

Figures \ref{mdpdpst_mag_rhb0_eta0_MC} and \ref{mdpdpst_mag_rhb0_eta5_MC}
show the effects of the Dirac sea (DS) as well
as the PV mixing on the masses of $D^+$ and ${D^*}^+$ mesons
 for the isospin symmetric and asymmetric 
(with $\eta$=0) nuclear matter
in (b) and (d) and compared with the case when the Dirac sea 
contributions are not taken into account (shown in (a) and (c) 
respectively). In the `no sea' approximation (Dirac sea neglected), 
the values of the scalar fields, and hence the open charm meson
masses calculated within the chiral effective model is observed
to be insensitive to the change in magnetic field
in the absence of the Landau level contributions \cite{dmeson_mag}.
The increase in the masses of $D^+$ and ${{D^*}^+}^{||}$ ($S_z$=0) 
are due to the positive Landau contributions 
in (a) for the case of 'no sea' approximation.
There is observed to be a drop (increase) in the mass
of the $D^+$ (${{D^*}^+}^{||}$) meson due to PV mixing.
The effect of inclusion of the AMMs of the nucleons 
is observed to be quite crucial on the masses of these mesons,
due to the fact that there is inverse magnetic catalysis,
as opposed to the magnetic catalysis observed when AMMs
are ignored. In the absence of AMMs of the nucleons, 
there is observed to be moderate increase in the masses 
$D^+$ and ${D^+}^*$ with magnetic field, whereas, for 
the case of accounting for the AMMs, there is observed
to be large drop in the masses at high values of the magnetic 
field. 

In figures \ref{md0d0st_mag_rhb0_eta0_MC} and
\ref{md0d0st_mag_rhb0_eta5_MC},
the effects of the Dirac sea (DS) contributions and PV
mixing are shown on the masses are plotted for $D^0$
and ${D^*}^0$ mesons for $\rho_B=\rho_0$
and for $\eta=0$ and $\eta=0.5$ respectively. 
Similar to the case of $D^+$ and $ {D^*}^+$,
the Dirac sea contributions have a significant effect
on the masses of these neutral open charm mesons.

The masses of the $\bar D$ and $\bar {D^*}$ mesons
are plotted for the charged sector
in figures \ref{mdmdmst_mag_rhb0_eta0_MC} and
\ref{mdmdmst_mag_rhb0_eta5_MC},
and for the neutral sector 
in figures \ref{md0bd0bst_mag_rhb0_eta0_MC} and
\ref{md0bd0bst_mag_rhb0_eta5_MC},
for $\eta=0$ and $\eta=0.5$
respectively. It is observed that the most dominant 
contribution due to the magnetic field 
on the masses of the open charm mesons
arise from the effects of the Dirac sea.

\section{Summary}
To summarize, we have investigated the effects of magnetic field
on the masses of the open charm pseudoscalar ($D$ and $\bar D$) 
and vector (${D^*}$ and $\bar {D^*}$) in magnetized (nuclear)
matter, accounting for the effects from the Dirac sea (DS), 
PV mixing and additional Landau level contribtions for 
the charged mesons. The effects of isospin asymmetry
as well as AMMs of the nucleons are also studied in the
present work. 
The effects of the Dirac sea is observed to lead to an 
enhancement (reduction) of the quark condensates, 
an effect called (inverse) magnetic catalysis,
which is observed as an increase (drop) in the magnitude of the scalar
fields, $\sigma$ and $\zeta$. At zero density, there is 
observed to be magnetic catalysis (MC) effect, which
is observed to be enhanced when 
the AMMs of the nucleons are also taken into account. 
At $\rho_B=\rho_0$, both for symmetric ($\eta$=0) and
asymmetric nuclear matter (with $\eta$=0.5), there 
is still observed to be magnetic catalysis when the
AMMs of the nucleons are neglected, whereas, the
trend becomes opposite with the magnetic field,
leading to inverse magnetic catalysis (IMC) in the
presence of the AMMs of the nucleons.
The effects of the Dirac sea contribution 
on the masses of the open charm mesons
is observed to be the most dominant effect
due to the magnetic field, the effect being much larger
than the PV mixing. The effect should modify the
yields pf the open and hidden charm mesons arising
from ultra-relativistic peripheral heavy ion
collision experiments, where the created magnetic field
is huge.

\end{document}